# Fast C-V method to mitigate effects of deep levels in CIGS doping profiles

P. K. Paul[1], J. Bailey[2], G. Zapalac[2], and A. R. Arehart[1]

[1]Electrical and Computer Engineering, The Ohio State University, Columbus, OH USA
[2]MiaSolé Hi-Tech Corp., Santa Clara, CA, USA

*Abstract* — In this work, methods to determine more accurate doping profiles in semiconductors is explored where trap-induced artifacts such as hysteresis and doping artifacts are observed. Specifically in CIGS, it is shown that this fast capacitance-voltage (C-V) approach presented here allows for accurate doping profile measurement even at room temperature, which is typically not possible due to the large ratio of trap concentration to doping. Using deep level transient spectroscopy (DLTS) measurement, the deep trap responsible for the abnormal C-V measurement above 200 K is identified. Importantly, this fast C-V can be used for fast evaluation on the production line to monitor the true doping concentration, and even estimate the trap concentration. Additionally, the influence of high conductance on the apparent doping profile at different temperature is investigated.

## INTRODUCTION

Accurate measurement of doping profiles is essential for accurate solar cell production, optimizing solar cell performance, and proper modeling and characterization. Typically, people use Hall measurement or capacitance-based approaches such as capacitance-voltage (C-V) or drive-level capacitance profiling (DLCP) to extract doping profiles [1-3]. However, the extracted doping can be influenced by deep levels and interface states[4-6]. In CIGS in particular, people use low temperature C-V or DLCP measurements. However, all of these techniques have their advantages and disadvantages [2,7,8]. For instance, Hall effect is a lateral technique, cannot be performed on actual solar cell structures [9]. The doping extracted from Hall measurement also does not provide any depth dependence, so non-uniformly doped samples can be problematic. Room temperature C-V measurements provides the depth dependant doping profile but the extracted doping profile can show large trap-induced hysteresis behavior [6,7]. Low temperature C-V measurements can successfully mitigate the hysteresis by freezing the effects of interface and bulk deep levels to measure the accurate doping concentration but it requires special equipment and longer total experiment times due to the cooling and heating [8]. Room temperature and low temperature DLCP measurements are used to eliminate the overestimation of doping concentration due to deep levels but DLCP measurements requires larger number of data acquisition and processing compared to C-V measurements and can not measure accurate doping profiles in non-uniform devices [10].

In CIGS solar cells, U-shaped doping profiles are commonly observed and suspected to be at least partially influenced by deep levels [7,8]. Some people use low temperature to slow down and avoid the trapping effects, and while this works it is more difficult and a priori knowledge of the defect time constants is required to ensure the trap emission is sufficiently slow at the measurement temperature [5]. Interpreting the actual doping profile from the temperature dependent extracted apparent U-shape doping profile, is a matter of debate [5,7]. Some groups consider the minimum point of the doping profile as the actual doping concentration [5] while others consider the highest reverse bias doping as the actual doping concentration [7]. Therefore, there is a strong need to understand the true doping profile.

In this paper, the deep trap responsible for the hysteresis in room temperature C-V is identified using deep level transient spectroscopy and a fast C-V measurement technique is proposed to avoid the influence of traps and accurately measure the doping profile even at room temperature. The trap-induced hysteresis and erroneous doping profiles is not limited to CIGS. Any semiconductor material system where the trap density is comparable to the doping density is subject to these issues, and the fast C-V approach is potentially applicable to all of these materials to achieve accurate doping profiles.

## APPROACH

In this study, CIGS solar cells were grown by a roll-to-roll sputter deposition process on a flexible stainless steel substrate by MiaSolé [11]. First, the Mo metal back contact was deposited on the steel substrate followed by the sputter deposition of the CIGS absorber layer. Finally, the CdS buffer layer and transparent conducting oxide window layer were deposited. Then Ni/Al/Ni Ohmic top contacts were evaporated on the aluminum doped zinc oxide (AZO) and the devices were physically circumscribed to isolate approximately 2 mm$^2$ devices.

The fast C-V measurements were performed with an Agilent function generator and Boonton 7200 capacitance meter with a 100 kHz bandwidth. A triangular voltage ramp was applied to the device with variable ramp rates, and the capacitance was measured with a 1 MHz 30 mVp-p AC signal. DC voltage from -1.0 and up to 0.3 V were used during C-V measurements, and the capacitance, conductance, and voltage were simultaneously recorded with a National Instruments data acquisition card. Net doping profiles (*N*) were calculated using [1]

$$N = \frac{-C^3}{q\varepsilon_s A^2 \left(\frac{dC}{dV}\right)} \qquad (1)$$

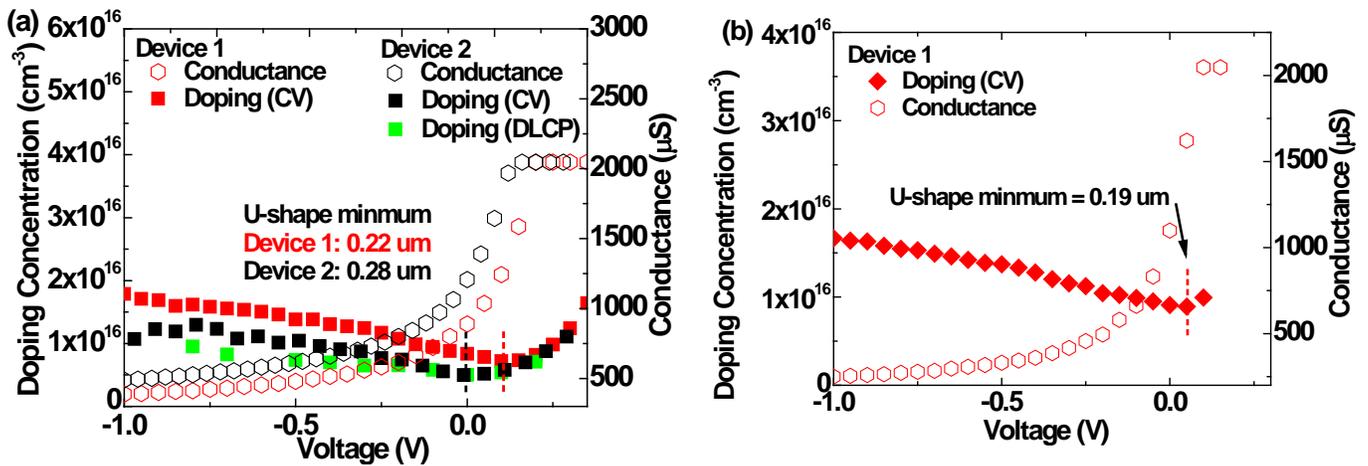

Figure 1: (a) Measured doping profiles by C-V using a 2 mV/s DC sweep rate and DLCP and conductance for two devices on the same sample at 110 K. The difference in devices is primarily in the amount of conductance. The DLCP and C-V extracted doping match quite well, but Devices 1 and 2 show different locations for the U-shaped doping, which likely arises in these devices because of the high conductance in forward bias causes artifacts in the measured capacitance – the doping begins to rise when the conductance approaches the maximum value of the meter (2047 µS). (b) Measured doping profiles by C-V and conductance for Device 1 at 210 K. the depletion depth at which the U-shaped minimum occurs has shifted with temperature suggesting it is not real. Like (a), the position of the U-shaped minimum is related to the conductance rise.

where $A$ is the device area, $\varepsilon_s$ is the permittivity, and $q$ is the elementary charge.

To characterize the deep traps, deep level transient spectroscopy (DLTS) was performed from temperature range 80K to 325 K [12-14]. The DLTS transients were analyzed using the double boxcar method with rate windows from 0.8 to 2000 s$^{-1}$. For the DLTS measurement, traps were filled with a +0.2 V pulse for 10 ms and the trap emission was recorded in reverse bias with a -1.0 V applied. The activation energy of the deep trap was determined by Arrhenius analysis and the concentration was determined from DLTS signal peak height accounting the lamda effect [1].

## RESULTS AND DISCUSSION

In Fig. 1, the CIGS solar cell doping profiles were extracted using C-V and DLCP measurements for two devices. These measurements were performed at 110 K and 210 K to avoid any hysteresis effects observed at higher temperatures. The extracted doping profile shows the typical U-shape doping concentration in the CIGS absorber layer. The depletion depth for minimum of the U-shape varies with both temperature and device suggesting it is the result of the measurement equipment namely the high conductance [6]. To explore this, the conductance was simultaneously recorded with the capacitance, which is also shown in Fig. 1. In Fig. 1(a), the doping profiles for two devices from the same sample were measured where the only difference was nominally the magnitude of the conductance. Both C-V and DLCP show very similar trends for the U-shaped minimum. However, the depletion depth at which the minimum occurs shifts from 0.22 to 0.28 µm as the conductance increases, which is likely because the high conductance in forward bias will at some point corrupt the measured capacitance signal leading to erroneous extract doping profile.

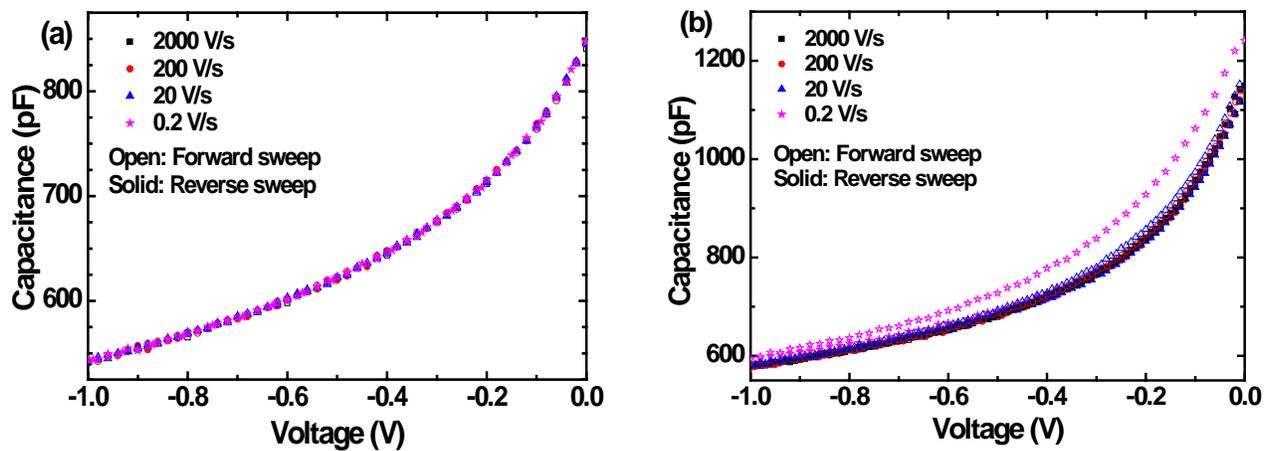

Figure 2: Measured C-V data measured at (a) 110 K and (b) 270 K with sweep rates of 2000 V/s, 200 V/s, 20V/s and 0.2 V/s where the reverse sweep was first followed by the forward sweep with no delay between the two directions. At 110 K no hysteresis is observed in the C-V measurements but at 270 K a signficant hysteresis is observed especially at the lowest sweep rate of 0.2 V/s.

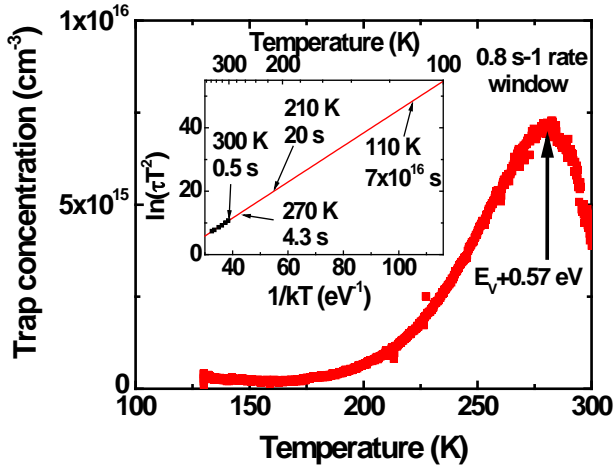

Figure 3: DLTS measurement on CIGS solar cell. DLTS spectra shows one dominant deep trap with activation energy $E_V+0.57$ eV. The inset shows the Arrhenius plot of the $E_V+0.57$ eV trap with estimated time constants for the various temperatures of measurement. The $E_V+0.57$ eV trap emission time constant $\tau$ is 0.5 s at 300 K, 4.3 s at 270 K, 20 s at 210 K, and $7 \times 10^{16}$ s at 110 K. This confirms the traps are frozen at 110 K and can respond quickly to the slow C-V sweep rate at 300 K.

To further explore this, Device 1 was measured at 210 K in Fig. 1(b) where the U-shaped minimum is reduced 0.03 μm compared with the 110 K case indicating again that conductance is likely playing a role in forward bias, but further work is needed to confirm this. Typically for high-fidelity capacitance measurements, it is desirable to have a Q factor ($Q=\omega C_P/G$ where $\omega$ is the AC angular velocity, $C_P$ is the parallel capacitance, and $G$ is the conductance) of 10, so as G increases Q decreases and the accuracy of the capacitance measurement decreases to the point where it cannot be trusted ($Q < \sim 1$) without additional verification. Forward bias is avoided in the rest of the C-V measurements to avoid these possible artifacts.

Usually, above 210 K the C-V measurements start to show hysteresis where the forward and reverse sweep capacitance do not match [8]. To understand how the hysteresis was affected by temperature and DC sweep rate, C-V measurements with forward and reverse sweeps were performed with different sweep rates at 110 K and 270 K, and the results are shown in Fig. 2. The 110 K C-V measurements in Fig. 2(a) show no distinguishable hysteresis at any sweep rate while the 270 K C-V measurements in Fig. 2(b) also show no hysteresis for DC sweep rates down to 20 V/s but shows large hysteresis of up to 100 pF at 0.2 V/s sweep rate. Traps were the most likely source of the hysteresis and the trap time constant depends exponentially on temperature. Therefore, defect spectroscopy was performed to identify the time constants and concentration of the deep levels.

Fig. 3 shows the DLTS spectra with one dominant trap with $E_V+0.57$ eV activation energy and a minimum trap concentration of $7 \times 10^{15}$ cm$^{-3}$. Previously, with scanning-DLTS the $E_V+0.57$ eV was found to be spatially localized and located only in specific intergrain regions [15-17]. The inset of Fig. 3 shows the Arrhenius plot of $E_V+0.57$ eV trap and the estimated trap emission time constants at several temperatures. The trap time constant $\tau_p$ follows [1]

$$\tau_p = \frac{1}{\sigma_p \langle v_{th} \rangle N_V} \exp\left(\frac{E_T-E_V}{kT}\right) \quad (2)$$

where $\sigma_p$ is the hole capture cross section, $N_v$ is the valence band density of states, $v_{th}$ is the thermal velocity, $E_V - E_T$ is the trap energy relative to the valence band, $T$ is the temperature, and $k$ is Bolzmann's constant. The trap emission time constants were compared to the total C-V measurement time to determine if the traps had time to emit during the measurement. The total C-V measurement times were 1 ms, 10 ms, 100 ms, 10 s for the 2000, 200, 20 and 0.2 V/s sweep rates, respectively. At 110 K (Fig. 2a), the $E_V+0.57$ eV trap emission time constant is several orders of magnitude larger than the measurement time for all cases, so the traps could not response to the DC bias change and hence there was no observed hysteresis. In contrast, at 270 K (Fig. 2(b)) the trap emission time constant is 4.3 s so the 0.2 V/s sweep (10 s total measurement time) is longer then the trap time constant and therefore the trap cause hysteresis, which is also because the trap density is comparable to the doping density. Still, the 2000, 200, and 20 V/s sweep rates at 270 K were much faster than the trap time constant (4.3 s), so the traps could not respond and hence no hysteresis was observed  Hence, we can conclude that the hysteresis behavior observed in the C-V is due to the $E_V+0.57$ eV trap.

Knowing the trap emission time constants from the Arrhenius plot, it is then possible to design C-Vs with fast and slow sweep rates as in Fig. 4, and then both the doping density and trap density can be measured. The total measurement time for the forward and reverse sweep $t_m$ is,

$$t_m = \frac{2\Delta V}{r_{DC}} \quad (3)$$

where, $r_{DC}$ is the DC bias sweep rate and $\Delta V$ is the measurement voltage. The maximum $r_{DC}$ is limited by the capacitance meter bandwidth ($BW$) and is

$$r_{DC,max} = V_{res}BW \tag{4}$$

where, $V_{res}$ is the minimum voltage resolution and 20 mV was assumed here. With a bandwidth of 100 kHz. this gives an $r_{DC,max}$ of 2000 V/s for this setup. The maximum sweep rate can be experimentally confirmed as well by increasing $r_{DC}$ until the C-V hysteresis begins to increase indicating a BW limited measurement. The minimum sweep rate ($r_{DC,min}$) is determined by the trap time constant. The total measurement time should be less than 10% of the trap time constant, so that most of the traps cannot emit during the measurement. This gives

$$r_{DC,min} = \frac{20\Delta V}{\tau_p} \tag{5}$$

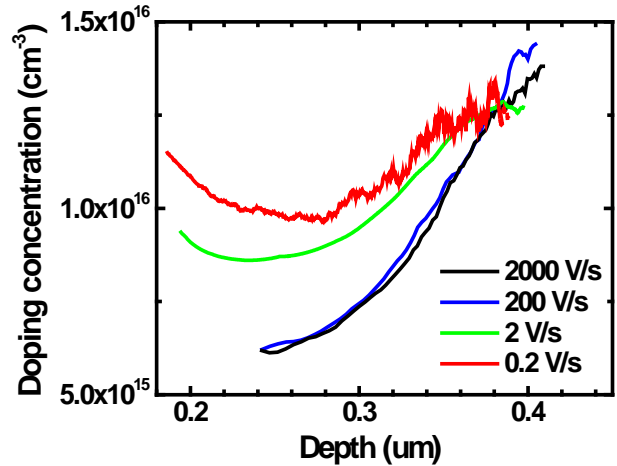

**Figure 4:** Measured doping profiles at 300 K for sweep rates listed in the legend.

For the $E_v$+0.57 eV trap at 300 K, $r_{DC,min}$ is 40 V/s. For $r_{DC}$ lower than 40 V/s, the trap-induced hysteresis would become visible and for $r_{DC}$ higher than this the hysteresis would be negligible. Therefore, above 40 V/s traps cannot respond and the doping profiles will exclude any trapping effects (i.e. is only the doping) and would then overlap at all temperatures, which is shown experimentally in Fig. 4 confirming the theory. Finally, the frequency of the AC test bias, which measures the out-of-phase current to calculate the capacitance, should be much larger than $1/\tau_p$. Here, the AC frequency used is 1 MHz which is well above the minimum required frequency. By meeting all these requirements, a high-fidelity C-V measurement can be achieved and a doping profile without the effects of trapping can be measured.

Additionally, it is possible with a fast and slow C-V to estimate the trap concentration. Using an $r_{DC}$ less than $0.2\Delta V/\tau_p$ (i.e. 0.4 V/s at 300 K for the $E_V$+0.57 eV trap) will provide sufficiently slow rates such that the trap will stay in equilibrium with the applied bias, and the doping profile extracted will be the sum of the trap and doping concentrations. The fast C-V only measured the true doping profile, so the difference between the fast and slow C-V doping profiles is an estimate of the trap concentration. Comparing the 2000 V/s (fast) and 0.2 V/s (slow) doping profiles in Fig. 4, a trap density in the low- to mid-$10^{15}$ cm$^{-3}$ is estimated, which agrees well with the DLTS result of 7x10$^{15}$ cm$^{-3}$ in Fig. 3. So, with a much faster and simpler measurement (fast and slow C-V) it is possible to get a quick and reasonably accurate estimates of both the trap concentration and doping profile at room temperature.

Finally, Fig. 5 shows the doping profile extracted from fast C-V using the highest 2000 V/s sweep rate for temperatures from 110 to 300 K. The extracted doping profiles are nominally identical over the whole temperature range indicating that traps are not influencing the doping profile at any of these temperatures. This indicates that the fast C-V approach to extract accurate doping profiles for CIGS or other material systems where the trap density is comparable with the doping density given the proper AC frequency and DC sweep rate. This also avoids the problem of needing to cool down thereby allowing for a simpler and cheaper setup to measure doping profiles.

## CONCLUSIONS

Using this fast C-V method, more accurate doping profiles can be obtained at room temperature. The U-shaped doping profiles in these samples is observed using C-V and DLCP when the conductance reaches high values. These results suggest high device conductance negatively influences the accuracy of the extracted doping profile. It is demonstrated that comparing the high and low sweep rate doping profiles that the difference is comparable with the measured $E_V$+0.57 eV trap density suggesting this trap is primarily responsible for the difference in measured doping, and the time constants of the trap and total measurement time can explain the onset of the observed hysteresis. Finally, this simple method can be extended for quick measurements during production to monitor defect densities and doping.

**Acknowledgements**
The authors would like to thank the Department of Energy (Contract #DE-DD0007141) for financial support.

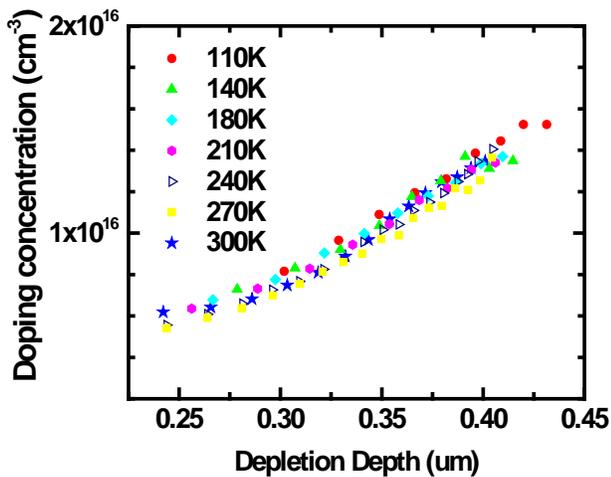

**Figure 5:** Measured doping profiles at different temperature with 2000 V/s sweep rate .